\documentclass[dvips]{acta}
\usepackage{supertabular,lscape,epsfig}
\usepackage{amssymb}
\usepackage{amsmath}
\SetPages{1}{5} 
\SetVol{62}{2012}

\begin{document}

\begin{Titlepage}

\Title { Stream Overflow in Z Cha and OY Car during Quiescence }

\Author {J.~~S m a k}
{N. Copernicus Astronomical Center, Polish Academy of Sciences,\\
Bartycka 18, 00-716 Warsaw, Poland\\
e-mail: jis@camk.edu.pl }

\Received{  }

\end{Titlepage}

\Abstract {
Eclipses of the hot spot in Z Cha and OY Car observed by many authors during their 
quiescence are re-analyzed. Distances of the spot from the center of the disk 
$r_s$ are determined from phases of ingress and egress.  
In the case of several eclipses of Z Cha and nearly all eclipses of OY Car it is found 
that $r_s{\rm (egress)}<r_s{\rm (ingress)}$. This implies that they are not 
representative for the radius of the disk $r_d$ and is interpreted as being due to 
the stream overflow. The $r_s({\rm ingress}) - dt$ relations (where $dt$ is 
the time since last outburst) are improved when points affected by the stream 
overflow are omitted. 
}
{binaries: cataclysmic variables, stars: dwarf novae, stars: individual: 
Z Cha, OY Car }

\section { Introduction } 

Hot spot eclipses observed in dwarf novae provide an important 
tool for determining dimensions of their accretion disks and studying 
their evolution during the dwarf nova cycle (see Warner 2003, Smak 1996a 
and references therein). 

In the standard approach it is assumed that the hot spot is produced 
by the collision of the stream with the outer parts of the disk. 
If so, the distance of the spot from the center of the disk $r_s$ is identical with 
the radius of the disk $r_d$. 
In the case of such a "standard" hot spot the phases of mid-ingress $\phi_i$ 
and mid-egress $\phi_e$ are related to the mass ratio (defining the shape 
of the stream trajectory), the orbital inclination, the radius of the disk, 
and a certain parameter $\Delta s$ describing the elongation of the spot 
along the disk's circumference (cf. Smak 1996a,2007). 
Conversely, when $q$, $i$, and $\Delta s$ are known, 
the phases of ingress and egres, $\phi_i$ and $\phi_e$, can be used to obtain 
two indpendent values of the spot distance: $r_s{\rm (ingress)}$ and 
$r_s{\rm (egress)}$. It is obvious that they should be identical and this 
provides the crucial test for the applicability of the concept of a "standard" 
hot spot and -- in particular -- of the assumption that the radial distance 
of the spot is identical with the radius of the disk. 

From the analysis of spot eclipses in U Gem (Smak 1971, 1996a) and IP Peg 
(Wolf et al, 1993, Wood et al. 1989b, Smak 1996a) it was found that the disk 
expands during normal outbursts and then slowly contracts during quiescence. 
This provides a strong observational support to the theory and models of dwarf 
nova outbursts (see Lasota 2001 and references therein). 

In the case of Z Cha, however, it was found (O'Donoghue 1986, Zo{\l}a 1989) 
that the radius of its disk also decreases during quiescence but the $r_d-dt$ 
diagram (where $dt$ is the time since the last outburst) is more poorly defined, 
showing large scatter of points. (Somewhat surprisingly no such scatter is 
present in the $r_d-\phi_{cycle}$ diagram, where $\phi_{cycle}$ is the phase 
of the outburst cycle). 
In the case of OY Car the situation was even more disappointing: the $r_d~-~dt$ 
diagram (and also the $r_d-\phi_{cycle}$ diagram) shows large scatter of points 
with no clear dependence of radius on $dt$ (cf. Smak 1996b). 

The assumption of a "standard" hot spot breaks down when the stream overflows the 
disk and part of its kinetic energy is dissipated along its trajectory above and below 
the surface of the disk. The hot spot is then replaced with a "hot stripe",  
its "effective distance" from the center, as determined from the eclipse analysis, 
being no longer representative for the radius of the disk. 
Such a situation was identified recently among some of the eclipses observed in 
Z Cha and OY Car during their superoutbursts (Smak 2007, 2009). 
As a characteristic feature of such eclipses it was found that the spot distance  
$r_s{\rm (egress)}$ determined from egress is smaller than the spot distance 
$r_s{\rm (ingress)}$ determined from ingress. 
This was interpreted as being due to selfabsorption effects 
in the "hot stripe" when it is observed during ingress; due to those effects the 
distances obtained from ingress, $r_s({\rm ingress})$, come out much closer to the true radii 
of the disk.  

The purpose of the present paper is to present similar evidence for Z Cha and OY Car 
at their quiescence, suggesting that their peculiarities discussed above 
can be explained as being due to the stream overflow.

\section { Z Cha } 

Eclipses in Z Cha at quiescence were observed extensively by Cook (1985b), 
Cook and Warner (1984), O'Donoghue (1986) and Wood et al. (1986). 
Using the phases of ingress and egress, $\phi_i$ and $\phi_e$, listed in those papers,  
with $q=0.20$, $i=80.2$ (Smak 2007) and $\Delta s=0.02$ (Smak 1996a),   
we obtain the spot distances $r_s{\rm (ingress)}$ and $r_s{\rm (egress)}$, 
which are compared in Fig.1. 

Most points do not deviate from the $r_s{\rm (egress)}=r_s{\rm (ingress)}$ 
line by more than 10 percent and there are no points with 
$r_s{\rm (egress)}>1.1 \times r_s{\rm (ingress)}$. On the other hand, however, 
there are several points with $r_s{\rm (egress)}<0.9 \times r_s{\rm (ingress)}$, 
suggesting that they represent the case of stream overflow. 

Supporting this interpretation is Fig.2, showing two $r_s-dt$ diagrams with values 
of $r_s$ obtained from ingress and egress. As expected, the $r_s{\rm (ingress)}-dt$ 
relation is quite well defined while the $r_s{\rm (egress)}-dt$ relation shows 
large scatter due to effects of stream overflow.

\begin{figure}[htb]
\epsfysize=4.5cm 
\hspace{4.0cm}
\epsfbox{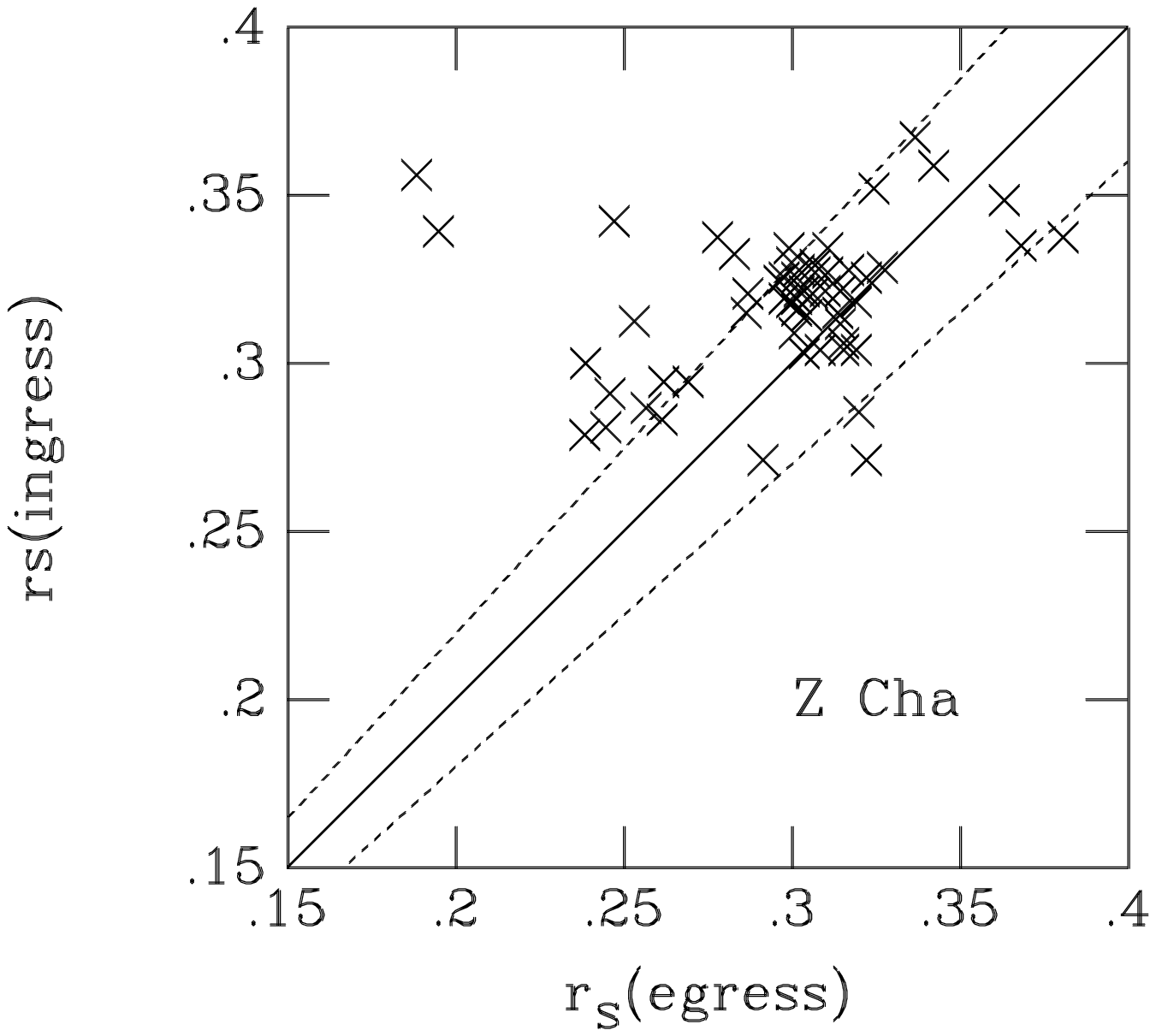} 
\vskip 5truemm
\FigCap { 
Comparison of $r_s{\rm (ingress)}$ and $r_s{\rm (egress)}$ obtained 
for Z Cha at quiescence. 
Dotted lines represent $r_s{\rm (egress)}= 0.9 \times r_s{\rm (ingress)}$ and 
$r_s{\rm (egress)}=1.1 \times r_s{\rm (ingress)}$. 
}
\end{figure}

\begin{figure}[htb]
\epsfysize=7.0cm 
\hspace{3.0cm}
\epsfbox{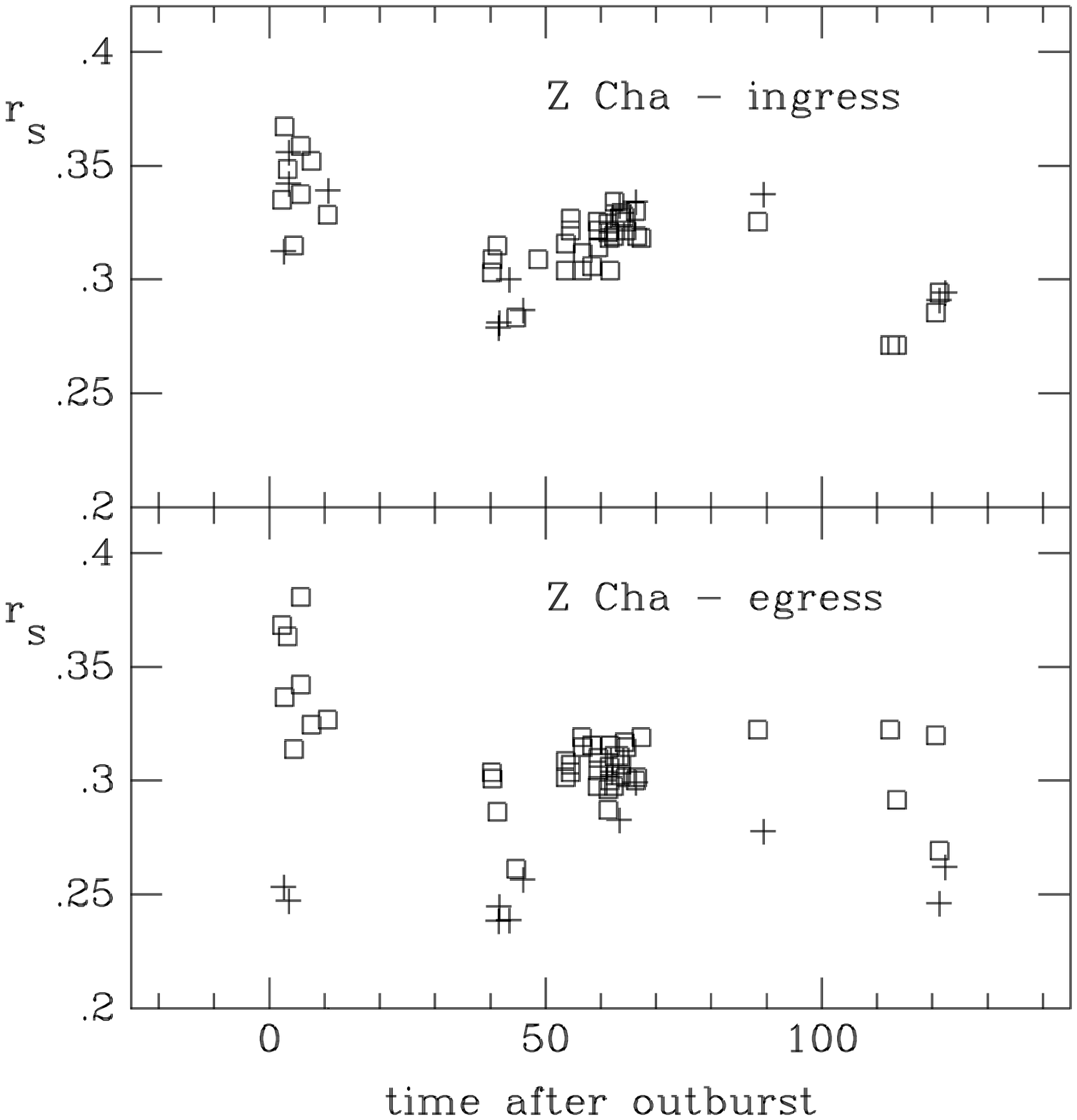} 
\vskip 5truemm
\FigCap { Spot distances in Z Cha at quiescence -- $r_s{\rm (ingress)}$ and 
$r_s{\rm (egress)}$ -- are plotted versus time (in days) since the last outburst. 
Points with $r_s{\rm (egress)}<0.9 \times r_s{\rm (ingress)}$, indicating 
stream overflow, are plotted as crosses. }
\end{figure}

\section { OY Car } 

OY Car and its eclipses were observed photometrically at quiescence by
Berriman (1984), Cook (1985a), Schoembs (1986), Schoembs and Hartmann (1983), 
Schoembs et al. (1987), Vogt (1983), Vogt et al. (1981) and Wood et al. (1989a). 
Using the phases of ingress and egress, $\phi_i$ and $\phi_e$, listed in their papers, 
with $q=0.10$, $i=83.3$ (Wood et al. 1989a) and $\Delta s=0.02$ (Smak 1996a),  
we obtain the spot distances $r_s{\rm (ingress)}$ and $r_s{\rm (egress)}$, 
which are compared in Fig.3. 

Practically all points in this diagram show $r_s{\rm (egress)}<r_s{\rm (ingress)}$, 
most of them deviating from the $r_s{\rm (ingress)}=r_s{\rm (egress)}$ line 
by more than 10 percent. This suggests that we are dealing with effects 
of the stream overflow.

\begin{figure}[htb]
\epsfysize=4.5cm 
\hspace{4.0cm}
\epsfbox{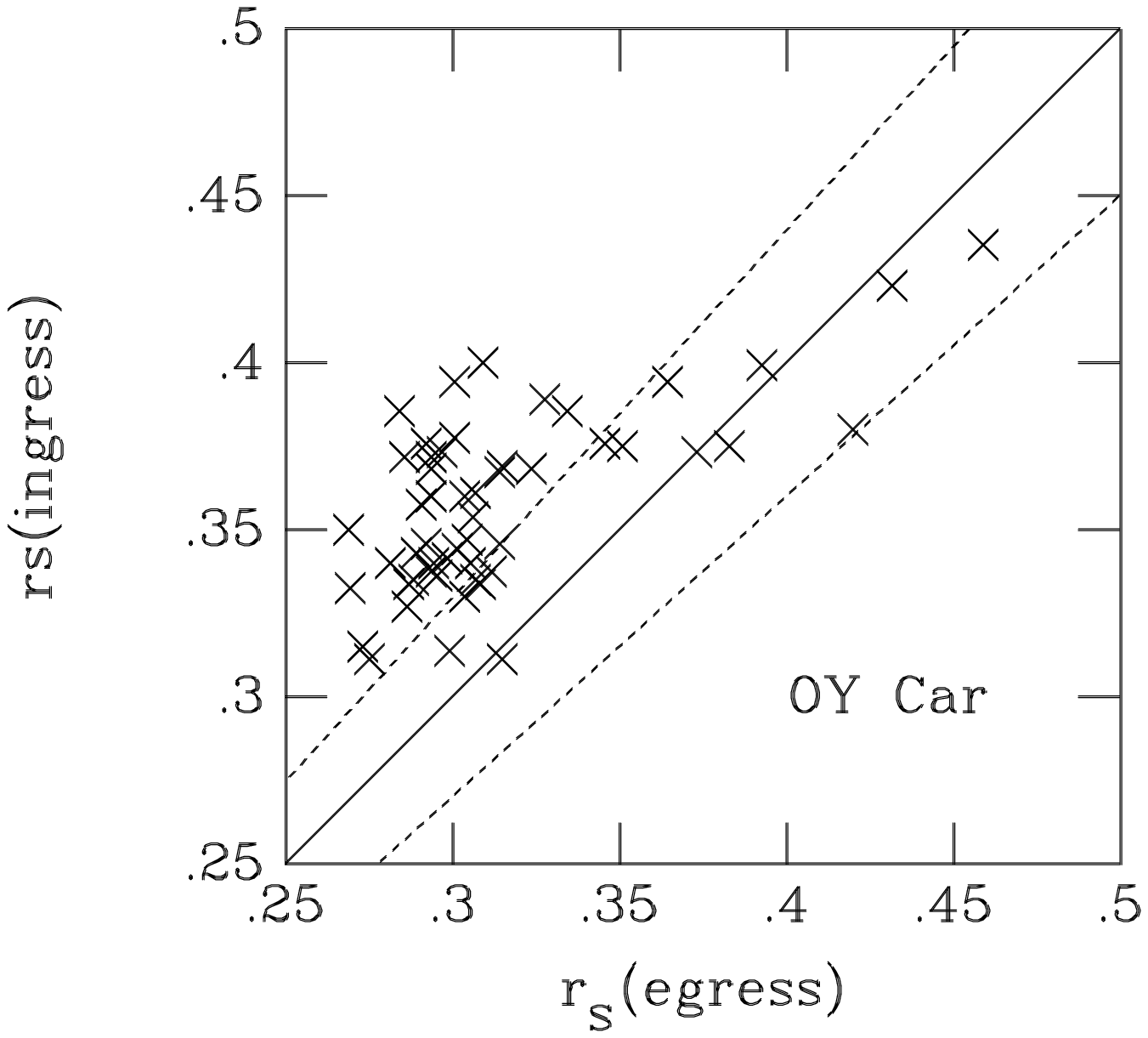} 
\vskip 5truemm
\FigCap { 
Comparison of $r_s{\rm (ingress)}$ and $r_s{\rm (egress)}$ obtained for OY Car 
at quiescence. 
Dotted lines represent $r_s{\rm (egress)}= 0.9 \times r_s{\rm (ingress)}$ and 
$r_s{\rm (egress)}=1.1 \times r_s{\rm (ingress)}$. }
\end{figure}

\begin{figure}[htb]
\epsfysize=7.0cm 
\hspace{3.0cm}
\epsfbox{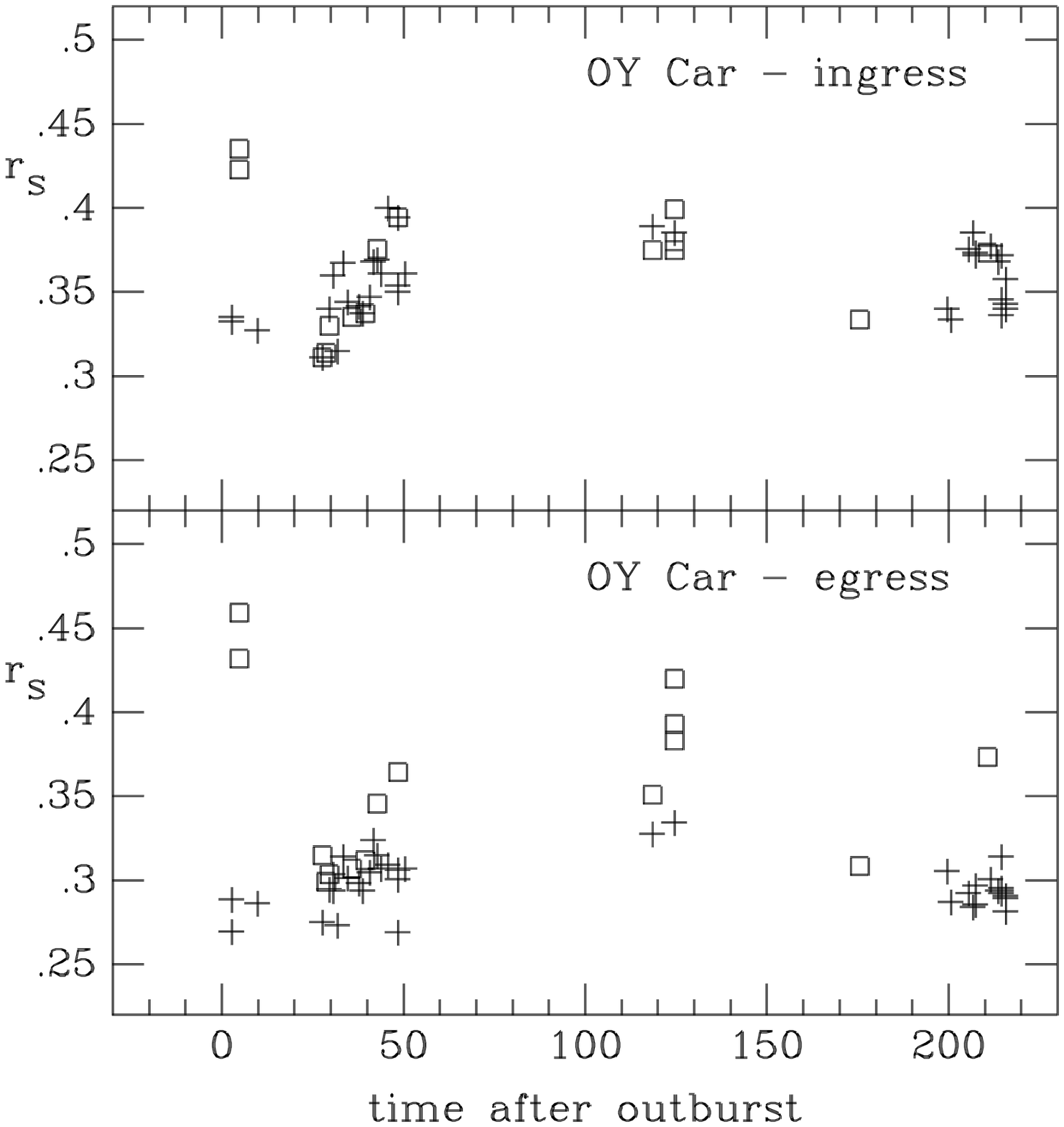} 
\vskip 5truemm
\FigCap { Spot distances in OY Car at quiescence -- $r_s{\rm (ingress)}$ and 
$r_s{\rm (egress)}$ -- are plotted versus time (in days) since the last outburst. 
Points with $r_s{\rm (egress)}<0.9 \times r_s{\rm (ingress)}$, indicating 
stream overflow, are plotted as crosses. 
}
\end{figure}

Shown in Fig.4 are the two $r_s-dt$ diagrams with values of $r_s$ obtained 
from ingress and egress. As expected, the $r_s{\rm (ingress)}-dt$ relation 
is defined much better than the $r_s{\rm (egress)}-dt$ relation. 
Note, however, that there are few squares (around $dt=40$ d), which deviate 
{\it in minus} from the $r_s{\rm (ingress)}-dt$ relation defined by other points 
with $r_s{\rm (ingress)}\approx r_s{\rm (egress)}$. 
This shows that -- when the stream overflow is present -- the equality 
$r_s{\rm (ingress)}=r_s{\rm (egress)}$ does not always imply $r_d=r_s$.

\section { Conclusion } 

It was predicted earlier by Hessman (1999) that stream overflow should occur most 
easily in quiescent dwarf novae, particularly those of the shortest orbital 
periods. Results for Z Cha and OY Car presented in this paper confirm 
those predictions. At the same time they form another important lesson, or warning: 
no reliable values of disk radii can be obtained from the analysis of eclipses 
of "hot stripes" produced by stream overflow.

\begin {references} 

\refitem {Berriman, G.} {1984} {\MNRAS} {207} {783} 

\refitem {Cook, M.C.} {1985a} {\MNRAS} {215} {211}

\refitem {Cook, M.C.} {1985b} {\MNRAS} {216} {219} 

\refitem {Cook, M.C. and Warner, B.} {1984} {\MNRAS} {207} {705} 

\refitem {Hessman, F.V.} {1999} {\ApJ} {510} {867} 

\refitem {Lasota, J.-P.} {2001} {New Astronomy Reviews} {45} {449}

\refitem {O'Donoghue, D.} {1986} {\MNRAS} {220} {23P} 

\refitem {Schoembs, R.} {1986} {\AA} {158} {233} 

\refitem {Schoembs, R. and Hartmann, K.} {1983} {\AA} {128} {37} 

\refitem {Schoembs, R., Dreier, H., and Barwig, H.} {1987} {\AA} {181} {50} 

\refitem {Smak, J.} {1971} {\Acta} {21} {15} 

\refitem {Smak, J.} {1996a} {\Acta} {46} {377} 

\refitem {Smak, J.} {1996b} {{\rm In } Cataclysmic Variables and Related Objects, 
         {\rm ed. J.H.Wood, Dordrecht: Kluwer}} {~} {p.45} 

\refitem {Smak, J.} {2007} {\Acta} {57} {87}

\refitem {Smak, J.} {2009} {\Acta} {59} {89}

\refitem {Vogt, N.} {1983} {\AA} {128} {29}

\refitem {Vogt, N., Schoembs, R., Krzemi{\'n}ski, W., and Pedersen, H.} 
         {1981} {\AA} {94} {L29}

\refitem {Warner, B.} {2003} { {\it Cataclysmic Variable Stars}, 
                      {\rm 2nd edition, Cambridge University Press } } {~} {~}

\refitem {Wolf, S., Mantel, K.H., Horne, K., Barwig, H., Schoembs, R. and 
          Baernbantner, O. } {1993} {\ApJ} {273} {160} 

\refitem {Wood, J.H., Horne, K., Berriman, G., O'Donoghue, D., and Warner, B.} 
         {1986} {\MNRAS} {219} {629}

\refitem {Wood, J.H., Horne, K., Berriman, G., and Wade, R.A.} 
         {1989a} {\ApJ} {341} {974}

\refitem {Wood, J.H., Marsh, T.R., Robinson, E.L., Stiening, R.F., Horne, K., 
          Stover, R.J., Schoembs, R., Allen, S.L., Bond, H.E., Jones, D.H.P., 
          Grauer, A.D., and Ciardullo, R. } 
          {1989b} {\MNRAS} {239} {809}

\refitem {Zo{\l}a, S.} {1989} {\Acta} {39} {45}

\end {references}

\end{document}